\documentclass[sigconf]{acmart}
\usepackage{eso-pic}

\AddToShipoutPictureBG*{%
  \ifnum\value{page}=1
    \put(50,770){\makebox[400pt][l]{\textbf{This paper has been accepted for publication on the 1st free5GC World Forum in conjunction with ACM CCS 2025}}}
  \fi
}

\AtBeginDocument{%
  }

\setcopyright{none}
\copyrightyear{2018}
\acmYear{2018}
\acmDOI{XXXXXXX.XXXXXXX}
\acmConference[Conference acronym 'XX]{Make sure to enter the correct
  conference title from your rights confirmation email}{June 03--05,
  2018}{Woodstock, NY}
\acmISBN{978-1-4503-XXXX-X/2018/06}

\usepackage{enumitem}

\begin{document}

\title{Robust Anomaly Detection in O-RAN: Leveraging LLMs against Data Manipulation Attacks}


\author{Thusitha Dayaratne}\affiliation{Monash University\country{Australia}}
\author{Ngoc Duy Pham}\affiliation{Monash University\country{Australia}}
\author{Viet Vo}\affiliation{Swinburne University of Technology\country{Australia}}
\author{Shangqi Lai}\affiliation{CSIRO's Data61\country{Australia}}
\author{Sharif Abuadbba}\affiliation{CSIRO's Data61\country{Australia}}
\author{Hajime Suzuki}\affiliation{CSIRO's Data61\country{Australia}}
\author{Xingliang Yuan}\affiliation{The University of Melbourne\country{Australia}}
\author{Carsten Rudolph}\affiliation{Monash University\country{Australia}}

\renewcommand{\shortauthors}{Dayaratne et al.}

\begin{abstract}
The introduction of 5G and the Open Radio Access Network (O-RAN) architecture has enabled more flexible and intelligent network deployments. However, the increased complexity and openness of these architectures also introduce novel security challenges, such as data manipulation attacks on the semi-standardised Shared Data Layer (SDL) within the O-RAN platform through malicious xApps. In particular, malicious xApps can exploit this vulnerability by introducing subtle Unicode-wise alterations (hypoglyphs) into the data that are being used by traditional machine learning (ML)-based anomaly detection methods. These Unicode-wise manipulations can potentially bypass detection and cause failures in anomaly detection systems based on traditional ML, such as AutoEncoders, which are unable to process hypoglyphed data without crashing. We investigate the use of Large Language Models (LLMs) for anomaly detection within the O-RAN architecture to address this challenge. We demonstrate that LLM-based xApps maintain robust operational performance and are capable of processing manipulated messages without crashing. While initial detection accuracy requires further improvements, our results highlight the robustness of LLMs to adversarial attacks such as hypoglyphs in input data. There is potential to use their adaptability through prompt engineering to further improve the accuracy, although this requires further research. Additionally, we show that LLMs achieve low detection latency (under 0.07 seconds), making them suitable for Near-Real-Time (Near-RT) RIC deployments.
\end{abstract}




\maketitle

\section{Introduction}
Deployments of 5G networks have significantly enhanced the consumer experience in mobile communications. Its' core capabilities, including enhanced Mobile Broadband (eMBB), massive Machine-Type Communications (mMTC), and Ultra-Reliable Low-Latency Communication (uRLLC), open immense opportunities across numerous sectors. However, the widespread deployment, architectural complexity, and safety-critical nature of these networks make them attractive targets for various adversaries, ranging from individual malicious actors to nation-states.

Several vulnerabilities exist in Layer-3 protocols, including the Radio Resource Control (RRC)\cite{ts38331}, which manages connection setup and bearer configuration, and the Non-Access Stratum (NAS)\cite{ts24501} that is responsible for mobility and session management between User Equipments (UEs) and core networks. Given the critical nature of these protocols and insufficient integrity protection/encryption in some messages~\cite{5gsepctor}, adversaries are attracted to these. Exploiting vulnerabilities of these protocols enables attacks such as Blind Denial of Service (DoS), Null cipher, Downlink DoS, and Lullaby attacks, causing session disruptions, battery drainage, or compromised security setups~\cite{5GReasoner,touchinguntouchables,adaptover}. Despite the security measures defined by the 3GPP, substantial enhancements to these protocols are challenging and often deferred to future standards given the backward compatibility issues. Thus, it is essential to detect these attacks as early as possible. However, detecting these attacks is challenging within traditional monolithic architectures given vendor-specific constraints.

In recent years, the Open Radio Access Network (O-RAN) architecture is emerging as a solution to enhance the controllability, security, and visibility of cellular networks by decoupling RAN hardware and software components and standardising interfaces. This decoupling allows network operators and third-party developers to implement flexible, scalable, and vendor-agnostic monitoring and response solutions. In particular, Near Real time RAN Intelligent Controller (Near-RT RIC) can be leveraged to inspect and control the control-plane messages with the use of Artificial Intelligence and Machine Learning (AI/ML) modules, often implemented as xApps (applications that runs on Near-RT RIC). Some recent studies have already explored the potential for both rule-based~\cite{5gsepctor} and traditional ML-based methods~\cite{detran,6gxsec} for Layer-3 attack detection within the O-RAN context.
Many existing anomaly detection methods deployed within O-RAN rely on shared data layers (SDL) to retrieve appropriate data. However, the SDL interface is not yet fully standardised given the novelty of O-RAN concept. Thus, this under standardised nature of the SDL creates a potential attack surface. In particular, malicious or compromised xApps could potentially manipulate or obfuscate message data. For example, manipulation can be subtle alterations to message names or parameters, such as using "hypoglyphs" (visually similar, distinct Unicode variations) to bypass traditional ML-based models~\cite{5gsepctor,6gxsec} as these models require precise patterns or statistical features for effective detection.

To address these critical vulnerabilities, we propose leveraging Large Language Models (LLMs) for anomaly detection in 5G Layer-3 protocols. In particular, we attempt to leverage LLMs as a more robust approach against data manipulation attacks originating from the SDL. Unlike traditional ML approaches that rely on syntactic patterns, LLMs possess capabilities to identify semantically equivalent inputs (despite syntactically manipulated), which makes LLMs resilient against data manipulation attacks, such as hypoglyphing. Our preliminary analysis depicts that LLMs based anomaly detection methods are resilient against such attacks and can complete the detection within O-RAN time constraints. To the best of our knowledge, this work is the first to specifically investigate the feasibility, robustness, and real-time practicality of LLM-based anomaly detection against data manipulation attacks within the SDL context of 5G O-RAN environments.

\section{System Overview}
\subsection{Threat Model}
While numerous attack vectors exist in the O-RAN context, including fake O-DU/O-RU nodes, fake UEs, and jammers, this work focuses on a highly plausible and sophisticated threat of malicious xApps deployed on the Near-RT RIC. Unlike more resource-intensive attacks (e.g., enrolling fake E2 nodes or jamming), deploying malicious xApps requires minimal financial investment, which makes it a significant threat.

Attackers can exploit the under standardised and evolving nature of the xApp platform to deploy such malicious xApps on the Near-RT RIC. Security vulnerabilities within the platform and lack of in-depth evaluation procedures could enable these deployments. Additionally, adversaries can exploit zero-day vulnerabilities similar to the other common well-established application platforms such as iOS and Android app stores despite existing security measures. We assume the malicious xApp possesses read/write access to the SDL. This assumption aligns with recent findings~\cite{wisec2024,hung2024security} and observations from the O-RAN Working Group 11 study on Near-RT RIC and xApp security~\cite{oran-wg11-security-ric-xapps}, which denote the absence of robust authentication or authorisation for databases, including the SDL, within the O-RAN ecosystem.

Our threat model specifically considers a malicious xApp exploiting its SDL access to manipulate data that is consumed by a xApps, which uses that data for anomaly detection based on traditional ML model. The adversary's objective is to evade detection by these existing ML models. However, instead of deleting on direct corruption of the data, we assume the malicious xApp introduces subtle Unicode-wise alteration ("hypoglyphs"—characters that are visually similar) to legitimate Layer-3 messages. For example, the malicious xApp can modify \textit{RRCSetupRequest} message to \textit{RRCSetupRequest} where despite looks exactly same, here Latin "C" (U+0043) was replaced with a Cyrillic "C" (U+0421), Latin "e" (U+0065) with a Cyrillic "e" (U+0435) and Latin "q" (U+0071) with a visually similar character (U+055B, Armenian small letter q).


\subsection{Framework}
\label{sec:framework}
\begin{figure}
    \centering
    \includegraphics[width=0.8\linewidth]{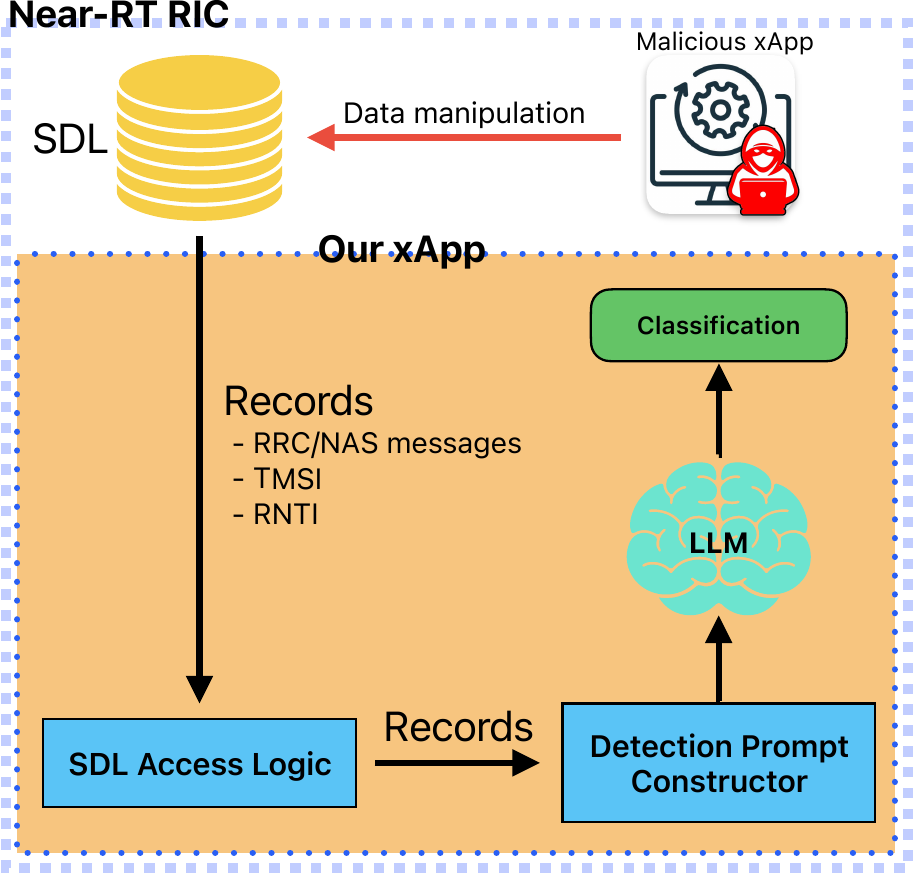}
    \caption{High-level system overview}
    \label{fig:framework}
\end{figure}

A high-level overview of the LLM-based detection framework is depicted in Figure~\ref{fig:framework}. The framework is deployed as an xApp within the Near-RT RIC and leverages the RRC and NAS messages, along with associated TMSI and RNTI identifiers that are stored in the SDL~\cite{5gsepctor}.

The SDL Access Logic module periodically queries the database to retrieve the latest RRC/NAS messages for attack detection. Upon retrieval, the messages are sent to Detection Prompt Constructor module. This module integrates the messages and a description about the task to construct the detection prompt, which is then use with the LLM to perform the classification. The LLM provides the classification (Normal or Anomalous).

\section{Implementation}
\subsection{Dataset}
We utilise the dataset from previous work~\cite{6gxsec}, which comprises data collected from a physical testbed involving four distinct 5G smartphone models. This dataset also incorporates data generated using the COLOSSEUM~\cite{colosseum} wireless network emulator. The dataset was structured in an ordered fashion, where it group messages sequentially by each UE (e.g., $msg_1^{UE_1}, msg_2^{UE_1}, \ldots, msg_1^{UE_2}, msg_2^{UE_2}, \ldots, \\msg_1^{UE_3}, msg_2^{UE_3}, \ldots$). However, the original dataset contains only three instances of Blind DoS attacks. Thus, we augmented the dataset by introducing 17 additional Blind DoS attack instances, which resulted about 1\% overall attack instances. In particular, we injected 17 \textit{RRCSetupRequest} messages with preselected existing TMSI values and random RNTI values. The enhanced dataset consists of 1,016 RRC and NAS messages, including 20 Blind DoS attack scenarios. 

Additionally, we further manipulated 5 messages within this enhanced dataset using Unicode-wise alteration, as discussed in our threat model in order to evaluate the resilience against evasion attacks. These manipulated messages include 2 of the Blind DoS attack instances and 3 normal messages. In particular, we changed some of the characters in those messages using alternative characters.

We then adopted a window-based approach with overlapping windows. Specifically, we empirically evaluated window sizes ranging from 1 to 10. A window size of 1 corresponds to the non-overlapping scenario, where each message is processed independently along with its corresponding previous message. In contrast, a window size of 10 indicates that each new incoming message is combined with the nine preceding messages. The corresponding previous message for the latest message is also included in the prompt. At any given time, we used only one new incoming message to enable streaming-based detection. This approach mitigates the inherent latency of batch processing, which occurs when multiple new messages are processed together. Such delays can allow malicious messages to remain undetected for extended periods and may require deferring processing until a predefined message threshold is reached.

\subsection{LLM}
\label{sec:llms}
We used the Meta's Llama-3.1-8B-Instruct model. The open-source model was obtained from HuggingFace\cite{huggingface} and deployed using vLLM\cite{vLLM} on an NVIDIA A100 GPU (80GB). We did not pre-train or fine-tune the model. The temperature parameter was set to 0 to ensure deterministic outputs and prevent LLMs from hallucinations or creative deviations in the predictions.

\section{Evaluation \& Results}
To evaluate the performance of the proposed framework, we aim to answer the following two research questions. 
\begin{enumerate}[label=\textbf{RQ-\arabic*}, leftmargin=*, labelwidth=3em, labelsep=1em, align=left]
    \item How does the presence of Unicode-wise manipulated messages (hypoglyphs) in the SDL impact the detection capabilities of traditional ML-based anomaly detection xApps?
    \item Does an LLM-based anomaly detection xApp demonstrate robustness against Unicode-wise manipulated (hypoglyphed) messages, and can its detection performance be effectively improved through prompt engineering?
\end{enumerate}

We primarily use Accuracy, Precision, Recall, F1 score, False Positive Rate (FPR), and False Negative Rate (FNR) for the evaluation. However, we note that achieving a high F1 Score along with low FPR and FNR is the desired outcome.

\subsection{RQ-1 Unicode-wise manipulated Layer-3 messages with traditional ML model}
Under this research question, we analysed the detection capabilities and robustness of existing traditional ML-based anomaly detection xApps when messages contain Unicode-wise manipulations (hypoglyphs). In particular, we aim to determine if these xApps can handle such adversarial attacks.

We utilised the AutoEncoder model and the code from the 6G-XSec~\cite{6gxsec}, an existing traditional ML-based anomaly detection xApp designed to identify Layer-3 attacks for this evaluation. We trained 10 distinct AutoEncoder models, where each model corresponding to a different window size ranging from 1 to 10. For each model, the first 700 normal messages from the dataset (with attacks filtered out) were used for training. The remaining data was used for testing.

\begin{figure}
    \centering
    \includegraphics[width=1\linewidth]{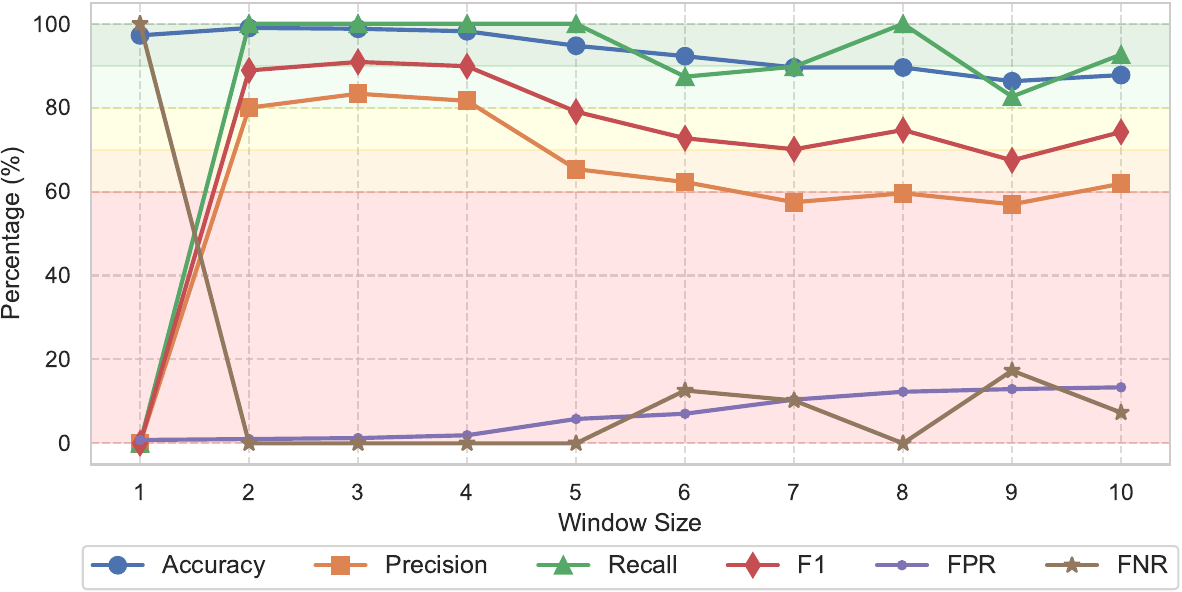}
    \caption{Detection performance of AutoEncoder model under normal scenario}
    \label{fig:ae}
\end{figure}

Figure \ref{fig:ae} depicts the detection performance of the models when tested without any hypoglyphed messages, where the F1 value remain over 75\% with most of the window sizes.

Then we tested the same models using the manipulated dataset, which includes 5 hypoglyphed messages. All 10 traditional ML-based AutoEncoder models experienced an immediate failure/crash upon the encounter of the first hypoglyphed message in the test set. The failure prevented any further anomaly detection. This failure is directly related to the AutoEncoder's underlying feature extraction and encoding mechanisms. In particular, the models were trained using a fixed set of normal, well-formed message patterns. Thus, they were unable to handle the novel, unseen Unicode characters introduced by the hypoglyphs. Unlike a misclassification, which would yield an incorrect output, the models' core processing pipelines failed. We denote that error handling could be implemented to prevent a full crash. However, that could potentially result in the model skipping or failing to process the un-encodable messages, which hinders the effectiveness of traditional ML-based anomaly detection against these evasion tactics.

\subsection{RQ-2 Effectiveness of LLM based Detection}
Under this research question, we primarily evaluated the robustness of an LLM-based anomaly detection xApp against hypoglyphed messages. Additionally, we also seek to analyse whether detection performance can be improved through prompt engineering. For this evaluation, we implemented an LLM-based anomaly detection xApp as discussed in section \ref{sec:framework}. The LLM-based xApp was evaluated using the complete dataset of 1,016 messages, which included the 5 hypoglyphed messages. Unlike in the AutoEncoder model, we initialised the same LLM model 10 times, corresponding to window sizes ranging from 1 to 10. 

Unlike the traditional ML-based AutoEncoder models, which experienced failures upon encountering hypoglyphed messages, LLM-based models showed the needed robustness. In particular, the LLM-based models successfully processed every single message in the dataset, including all Unicode-wise manipulated instances, without any system crashes or early terminations. This behaviour highlights that LLMs are not susceptible to the encoding and feature extraction failures, which hinder the impact of traditional models against unseen or subtly altered characters.

\begin{figure}
    \centering
    \includegraphics[width=1\linewidth]{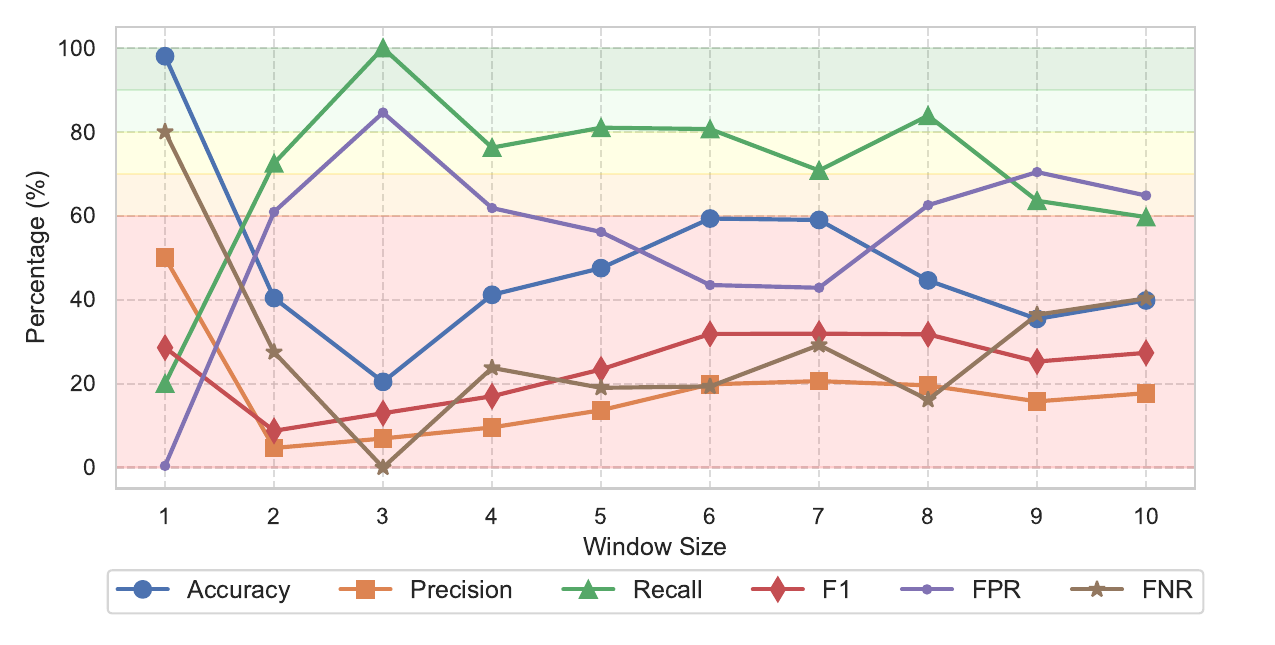}
    \caption{Detection performance of LLM-based models with hypoglyphed messages}
    \label{fig:llm}
\end{figure}

Figure \ref{fig:llm} depicts the detection performance of the LLM-based xApp across various window sizes.
The F1-scores ranged from about 0.087 to 0.319 across different window sizes. Despite the fact that the LLM successfully processed all inputs without crashing, these initial F1-scores suggest that the out-of-the-box anomaly detection capabilities for these Layer-3 attacks require further optimisation and prompt engineering.

Table \ref{tab:avg_time_window_size} shows the average detection time per message for each window size with the LLM-based approach. The average detection times ranged from 0.03 seconds for a window size of 1 to 0.07 seconds for a window size of 10. It is clear that the LLM requires more time when the number of messages in the window increases. However, these times are well within the Near-RT RIC's real-time constraint of 1 second, which confirms the practical feasibility of LLM-based detection in terms of latency. 

\begin{table}
\centering
\caption{Average detection time per message for LLM}
\vspace{-5mm}
\begin{tabular}{cc}
\toprule
\textbf{Window Size} & \textbf{Average Time (seconds)} \\
\midrule
1  & 0.030 \\
2  & 0.046 \\
3  & 0.056 \\
4  & 0.052 \\
5  & 0.054 \\
6  & 0.056 \\
7  & 0.056 \\
8  & 0.067 \\
9  & 0.069 \\
10 & 0.070 \\
\bottomrule
\end{tabular}
\label{tab:avg_time_window_size}
\end{table}

\section{Discussion \& Conclusion}
Our evaluation shows a critical issue in the pipeline of traditional ML-based anomaly detection methods in the O-RAN context, when exposed to semantically altered inputs, such as hypoglyphs. Instead of simple misclassifications, these inputs lead the models to operational failure upon processing manipulated data. This vulnerability is highly critical in the context of the O-RAN architecture, where a malicious xApp can exploit the evolving standardisation of the SDL to disable security monitoring systems without triggering conventional detection mechanisms. Such failures can potentially result in total loss of detection capabilities and leading the network vulnerable additional compromises.

In contrast, our LLM-based anomaly detection shows resiliency against manipulated (hypoglyphed) messages. This resilient nature of LLMs highlights LLM-based solutions as a reliable alternative to conventional ML models for security-critical deployments. Even though baseline detection performance requires additional optimisation, our findings show that LLMs can be used to implement anomaly detection solutions that are more robust against crashes from manipulated input data. We suggest that given the appropriate prompt engineering, their performance in terms of accuracy could be further improved without retraining. Additionally, our detection achieves below 0.1-second detection latency, which demonstrates its suitability for deployment in Near-RT RIC environments. These results demonstrate the potential of LLMs to provide robust, low-latency detection of Layer-3 attacks, even with advanced data manipulations within the SDL, contributing to the development of more secure and resilient O-RAN networks.

\section*{Acknowledgement}
This research paper is conducted under the 6G Security Research and Development Project, as led by the Commonwealth Scientific and Industrial Research Organisation (CSIRO) through funding appropriated by the Australian Government’s Department of Home Affairs. This paper does not reflect any Australian Government policy position. For more information regarding this Project, please refer to \url{https://research.csiro.au/6gsecurity/}.

\bibliographystyle{unsrtnat}
\bibliography{references}
\end{document}